\documentstyle[12pt]{article}
\def\PRL #1 #2 #3{{\sl Phys. Rev. Lett.} {\bf#1} (#2) #3}
\def\NPB #1 #2 #3{{\sl Nucl. Phys.} {\bf B#1} (#2) #3}
\def\NPBFS #1 #2 #3 #4{{\sl Nucl. Phys.} {\bf B#2} [FS#1] (#3) #4}
\def\CMP #1 #2 #3{{\sl Commun. Math. Phys.} {\bf #1} (#2) #3}
\def\PRD #1 #2 #3{{\sl Phys. Rev.} {\bf D#1} (#2) #3}
\def\PLA #1 #2 #3{{\sl Phys. Lett.} {\bf #1A} (#2) #3}
\def\PLB #1 #2 #3{{\sl Phys. Lett.} {\bf #1B} (#2) #3}
\def\JMP #1 #2 #3{{\sl J. Math. Phys.} {\bf #1} (#2) #3}
\def\PTP #1 #2 #3{{\sl Prog. Theor. Phys.} {\bf #1} (#2) #3}
\def\SPTP #1 #2 #3{{\sl Suppl. Prog. Theor. Phys.} {\bf #1} (#2) #3}
\def\AoP #1 #2 #3{{\sl Ann. of Phys.} {\bf #1} (#2) #3}
\def\PNAS #1 #2 #3{{\sl Proc. Natl. Acad. Sci. USA} {\bf #1} (#2) #3}
\def\RMP #1 #2 #3{{\sl Rev. Mod. Phys.} {\bf #1} (#2) #3}
\def\PR #1 #2 #3{{\sl Phys. Reports} {\bf #1} (#2) #3}
\def\AoM #1 #2 #3{{\sl Ann. of Math.} {\bf #1} (#2) #3}
\def\UMN #1 #2 #3{{\sl Usp. Mat. Nauk} {\bf #1} (#2) #3}
\def\FAP #1 #2 #3{{\sl Funkt. Anal. Prilozheniya} {\bf #1} (#2) #3}
\def\FAaIA #1 #2 #3{{\sl Functional Analysis and Its Application} {\bf
#1} (#2) #3}
\def\BAMS #1 #2 #3{{\sl Bull. Am. Math. Soc.} {\bf #1} (#2)
#3} \def\TAMS #1 #2 #3{{\sl Trans. Am. Math. Soc.} {\bf #1} (#2) #3}
\def\InvM #1 #2 #3{{\sl Invent. Math.} {\bf #1} (#2) #3}
\def\LMP #1 #2 #3{{\sl Letters in Math. Phys.} {\bf #1} (#2) #3}
\def\IJMPA #1 #2 #3{{\sl Int. J. Mod. Phys.} {\bf A#1} (#2) #3}
\def\AdM #1 #2 #3{{\sl Advances in Math.} {\bf #1} (#2) #3}
\def\RMaP #1 #2 #3{{\sl Reports on Math. Phys.} {\bf #1} (#2) #3}
\def\IJM #1 #2 #3{{\sl Ill. J. Math.} {\bf #1} (#2) #3}
\def\APP #1 #2 #3{{\sl Acta Phys. Polon.} {\bf #1} (#2) #3}
\def\TMP #1 #2 #3{{\sl Theor. Mat. Phys.} {\bf #1} (#2) #3}
\def\JPA #1 #2 #3{{\sl J. Physics} {\bf A#1} (#2) #3}
\def\JSM #1 #2 #3{{\sl J. Soviet Math.} {\bf #1} (#2) #3}
\def\MPLA #1 #2 #3{{\sl Mod. Phys. Lett.} {\bf A#1} (#2) #3}
\def\JETP #1 #2 #3{{\sl Sov. Phys. JETP} {\bf #1} (#2) #3}
\def\JETPL #1 #2 #3{{\sl  Sov. Phys. JETP Lett.} {\bf #1} (#2) #3}
\def\PHSA #1 #2 #3{{\sl Physica} {\bf A#1} (#2) #3}
\def\CQG #1 #2 #3{{\sl Class. Quantum Grav.} {\bf #1} (#2) #3}
\def\SJNP #1 #2 #3{{\sl Sov. J. Nucl. Phys. (Yadern.Fiz.)} {\bf #1} (#2) #3}
\def\a{\alpha}\def\b{\beta}\def\g{\gamma}\def\d{\delta}

\def\k{\kappa}\def\L{\Lambda}\def\s{\sigma}
\def\Th{\Theta}\def\G{\Gamma}
\def\u{\underline}
\def\be{\begin{equation}}\def\ee{\end{equation}}

\def\susy{supersymmetry}

\newcommand{\p}[1]{(\ref{#1})}
\begin{document}
\thispagestyle{empty}
\renewcommand{\thefootnote}{\fnsymbol{footnote}}
\begin{flushright}
hep-th/9711007\\
November 1997
\end{flushright}

\vspace{3truecm}
\begin{center}
{\large\bf Worldline Superfield Actions for N=2 Superparticles}

\vspace{1cm}Igor Bandos$^1$, Alexey Maznytsia$^2$
\footnote{e--mail: alex$\_{}$maznytsia@hotmail.com} and
Dmitri Sorokin$^1$\footnote{dsorokin@kipt.kharkov.ua}

\vspace{0.5cm}
$^1${\it National Science Center \\
Kharkov Institute of Physics and Technology, \\
Kharkov, 310108, Ukraine}

\bigskip
$^2${\it Department of Physics and Technology,
Kharkov State University \\
310108, Kharkov, Ukraine}

\vspace{1.cm}
{\bf Abstract}
\vspace{0.5cm}
\end{center}

We propose doubly supersymmetric actions in terms of $n=2(D-2)$ worldline
superfields for $N=2$ superparticles in $D=3,4$ and Type $IIA ~D=6$
superspaces. These actions are obtained by dimensional reduction of
superfield actions for $N=1$ superparticles in $D=4,6$ and $10$,
respectively. We show that in all these models geometrodynamical
constraints on target superspace coordinates do not put the theory on the
mass shell, so the actions constructed consistently describe the dynamics
of the corresponding $N=2$ superparticles.

We also find that in contrast to the $IIA ~D=6$ superparticle a chiral
$IIB ~D=6$ superparticle, which is not obtainable by dimensional reduction
from $N=1$, $D=10$, is described by superfield constraints which produce
dynamical equations. This implies that for the $IIB ~D=6$ superparticle
the doubly supersymmetric action does not exist in the conventional form.

\renewcommand{\thefootnote}{\arabic{footnote}}
\setcounter{footnote}0
\newpage
\section{Introduction}
An initial motivation for considering doubly supersymmetric models
\cite{dsusy}--\cite{hs1} was to better understand the relationship between
Ramond--Neveu--Schwarz and Green--Schwarz formulation of superstrings
\cite{gsw}. The former possesses manifest local supersymmetry on the
worldsheet of the superstring, while the latter is manifestly
supersymmetric in target superspace and, in addition, has a non--manifest
local fermionic (so called {\it kappa}) symmetry \cite{ks} on the
worldsheet. The doubly supersymmetric models possess both types of
supersymmetry simultaneously. In general they describe wider spectrum of
physical states than the single supersymmetric counterparts they stemmed
from \cite{dsusy}. At the same time in \cite{stv} it was realized that the
$\k$--symmetry of Casalbuoni--Brink--Schwarz superparticles \cite{cbs} and
Green--Schwarz superstrings \cite{gsw} can be a manifestation of a hidden
local supersymmetry on the world surface, and worldline superfield actions
for superparticles in $N=1,$ $D=3$ and $N=1$, $D=4$ target superspace were
constructed as an implementation of this idea. Since then all presently
known super--p--branes have acquired doubly supersymmetric description
(see \cite{bpstv} for a review and \cite{bpst9705} for recent progress)
thus forming a subclass of more general class of the doubly supersymmetric
models.

Having replaced the $\k$--symmetry with the local supersymmetry one got a
covariant algebra of irreducible first--class constraints generating this
symmetry, which has influenced the development of new methods of covariant
quantization of superparticles, superstrings \cite{ferber}--\cite{zima}
and null--super--$p$--branes \cite{zpbr}.

The doubly supersymmetric approach has also proved to be the most
appropriate for the application to studying super--p--brane dynamics of
geometrical methods of surface theory describing properties of embedding
world (super)surfaces into target (super)spaces \cite{bpstv,bsv,bpst9705}.
This has helped one to get superfield equations of motion for new
important types of super--p--branes, such as Dirichlet branes and a
five--brane of M theory \cite{hsw}, without knowledge of their actions,
which were constructed later on \cite{c,m}.

A basic condition which determines the embedding of world supersurface of
any super--p--brane into target superspace, is a so--called
geometrodynamical condition. It prescribes target--space supervielbein
vector components be zero along the Grassmann directions of the world
supersurface. Depending on the dimensions of worldvolume and target
superspaces this condition can define either non--minimal or minimal
embedding. For $N=1$ superparticles \cite{stv}--\cite{ds}, \cite{gs92},
Type I superstrings in $D=3,4,6$ and 10 \cite{hsstr}, and for an $N=2$
$D=3$ superparticle and superstring \cite{gs2} the  geometrodynamical
constraint defines non--minimal embedding. From the dynamical point of
view this means that the geometrodynamical constraint does not put these
 theories on the mass shell, i.e. it produces no dynamical equations of
motion. In this case this condition can be incorporated into the action
with a superfield Lagrange multiplier, and such an action will
consistently describe the dynamics of the supersymmetric object. All known
superfield actions of the models mentioned above contain this
geometrodynamical term.

However, for a wide class of models, such as Type II $D=10$ superstrings
and D--branes, the $N=1$, $D=11$ supermembrane and the five--brane the
situation is different. The geometrodynamical condition defines the
minimal embedding of the corresponding superworldvolumes into target
superspaces, i.e. it puts the theory on the mass shell
\cite{bpstv,bpst9705}. Now we cannot construct the geometrodynamical
action as in the case of non--minimal embedding since the superfield
Lagrange multiplier accompanying the geometrodynamical constraint  would
 contain redundant propagating degrees of freedom spoiling the spectrum of
physical states of the model. Thus, when the geometrodynamical constraint
 contains superfield equations of motion of a super--p--brane one
encounters a problem in constructing a worldvolume superfield action for
the super--p--brane\footnote{This problem of supersymmetric theories is
well known. When the number of supersymmetries is too large and/or the
dimension of space--time is too high, superfield constraints which are
required for diminishing a number of independent fields put the theory on
the mass shell.}.  In this case one can use a generalized action principle
of the group--manifold approach \cite{rheo}. It allows one to construct a
worldvolume functional which, however, is not a superfield action in the
conventional sense \cite{bsv}.

On the contrary, conventional superfield actions should exist for the
models with off--shell superfield constraints and it seems of interest to
find wider class of super--p--branes (than that known so far) for which
the off--shell geometrodynamical constraint does not produce equations of
 motion, and to construct superfield actions for them. A possible way of
getting, at least, some of these models is to consider the dimensional
reduction of super--p--brane models for which superfield actions are
already known \cite{stv}--\cite{gs2}. For instance, we can take an $N=1$,
$D=10$ superparticle \cite{gs92} and dimensionally reduce it down to
$N=2$, $D=6$, or reduce an $N=1$, $D=6$ superparticle to $N=2$, $D=4$ and
see what form of doubly supersymmetric actions one gets for these
superparticles with extended supersymmetries.  This is the main purpose of
this paper.

	The paper is organized as follows:

	In Section 2 we describe the main features of the doubly
supersymmetric formulation of an $N=1$ superparticle which will be then
used to obtain superfield actions of $N=2$ superparticles in $D=3,4$ and
$D=6$.

	In Section 3 the action of an  $N=2, ~D=3$ superparticle is
	obtained by the dimensional reduction of the superfield action of
the $N=1$ superparticle in $D=4$. In addition to the geometrodynamical
term (which was considered earlier by Galperin and Sokatchev \cite{gs2})
this action includes a Lagrange multiplier term with a bilinear
combination of Grassmann derivatives of the fermionic superfields. The
Lagrange multiplier is a purely auxiliary degree of freedom, so its
presence does not spoil the physical content of the model. The second term
in the action ensures its invariance under a local symmetry with
superfield parameters, analogous to that of $N=1$ superparticles
\cite{gs92}, which allows one to gauge away auxiliary fields from the
Lagrange multipliers.

       In Sections 4 and 5 we apply the dimensional reduction procedure to
get doubly supersymmetric actions for an $N=2$ superparticle in $D=4$ and
a Type $IIA$ superparticle in $D=6$. We analyze the geometrodynamical
 constraints in the both cases and show that they do not produce equations
of motion. We also demonstrate that for the Type $IIB$ $D=6$ superparticle
it turns out to be impossible to construct the superfield action in the
form mentioned above because in this case the geometrodynamical constraint
puts the theory on the mass shell. The dynamical contents of the models
constructed coincide with that of the $D=4$ and $6$ $N=2$
Casalbuoni--Brink--Schwarz superparticles.

{\it Conventions and notation}

    We use the mostly negative signature for the Minkowski metric tensor
$$
\eta ^{\u {mn}}=diag(+,-,...,-).
$$

    In what follows the indices corresponding to vector and spinor
representations of target space Lorentz groups $SO(1,D-1)$ are underlined;
non--underlined indices with and without hats are reserved for
representations of orthogonal and unitary groups, respectively.

    $SU(2)$--indices are raised and lowered by the unit antisymmetric
    tensors ${\epsilon}^{AB}$ and ${\epsilon}_{AB}$ \\
$({\epsilon}_{21}={\epsilon}^{12}=1)$ as follows $$
{\xi}^A={\epsilon}^{AB}{\xi}_B,~ {\xi}_A={\epsilon}_{AB}{\xi}^B.~ $$ $\{
...\} $ and $[...]$ denote symmetrization and antisymmetrization of
indices with the weight 1 respectively.  Other notation are introduced
below.

\section{$N=1$ superparticle in terms of unrestricted world\- line
superfields}

       In this section we briefly consider the doubly supersymmetric
formulation of $N=1$ superparticle mechanics in $D=3,4,6$ and $10$
\cite{stv,gs92} in terms of unrestricted worldline superfields.

       The superparticle dynamics is defined by the minimal embedding of
the one--dimensio\- nal $n=D-2$ worldline supersurface into the
$D$--dimensional $N=1$ target superspace \footnote{The number of worldline
supersymmetries is chosen to be equal to the number of independent
$\kappa$--symmetry transformations of the corresponding Brink--Schwarz
superparticle action, i.e. half of the number of target space
supersymmetries \cite{stv}.} (we will consider flat target superspaces
only). In the case of the doubly supersymmetric particles
\cite{stv}--\cite{gs92}, \cite{gs2} the intrinsic geometry of the
worldline is superconformally flat and can be described with the use of
the flat Cartan forms \cite{stv,ghs,gs92} \be \label{201} e_{\tau}=d\tau
-id\eta _{\hat q}\eta _{\hat q},\qquad e_{\hat q}=d\eta _{\hat q}, \ee
where ${\hat q}$ is the index of the representation of the $SO(D-2)$
group, which is the automorphism group of the $(1\vert D-2)$ worldline
supersymmetry algebra. These forms constitute a non--degenerate
supervielbein in the cotangent space of the worldline supermanifold. It is
used as a basis of the superworldline differential forms. For example,
pullbacks of superinvariant one--forms $\Pi ^{\u m}\equiv dX^{\u
m}-id{\bar \Th}{\G}^{\u m}{\Th}$ and $\Pi ^{\u \mu}\equiv d\Th ^{\u \mu}$
of the flat target superspace onto the superworldline are \be \label{202}
\Pi ^{\u m}= e_{\tau}\Pi_{\tau}^{\u m}+e_{\hat q}\Pi ^{\u m}_{{\hat q}},
\ee \be \label{203} \Pi ^{\u \mu}= e_{\tau}\Pi_{\tau}^{\u \mu}+e_{\hat
q}\Pi ^{\u \mu}_{   {\hat q}}.  \ee

      Embedding equations adjust the target space supervielbein to the
      worldline one in such a way that the worldline vector component of
${\Pi}^{\u m}$ is directed along the bosonic component of the
superworldline frame, the fermionic components of the target space
supervielbein lie along Grassmann directions of the worldline superspace
and all other (bosonic) components of the target supervielbein are
orthogonal to the worldline. In particular, projections of the form $\Pi
^{\u m}$ onto the intrinsic vielbein forms $e_{\hat q}$ vanish:  \be
\label{204} \Pi ^{\u m}_{   {\hat q}}= D_{{\hat q}}X^{\u m}-iD_{{\hat
q}}{\bar \Th}{\G}^{\u m}{\Th}=0, \ee where $D_{ {\hat
q}}={{\partial}\over{\partial \eta _{\hat q}}}+ i\eta _{\hat
q}{{\partial}\over{\partial \tau }}$ are flat fermionic covariant
derivatives of the worldline superspace and the superparticle target space
coordinates $X^{\u m}$ and ${\Th}^{\u \mu}$ are scalar unrestricted
worldline superfields.

       Equation \p{204} is called {\it the  geometrodynamical constraint
}.  Its left--hand side $\Pi ^{\u m}_{   {\hat q}}$ is incorporated into
superfield actions of superparticles with a superfield Lagrange multiplier
$P_{{\u m}{\hat q}}$:  \be \label{205} S_D^{N=1}=\int d\tau d^{D-2}\eta
P_{{\u m}{\hat q}}{\Pi _{   {\hat q}}^{\u m}}.  \ee The integration is
performed over the worldline supersurface, which is supposed  to be
non--degenerate. This means \cite{gs92} that the vector ${{{\partial}X^{\u
m}}\over{{\partial}{\tau}}}$ is non--vanishing and the matrix $D_{\hat
q}{\Th}^{\u \mu}$ is of the maximal rank \footnote{The requirement
\p{2055} is introduced to exclude a non--physical solution corresponding
to a particle ``frozen" into a point of the target superspace}:  \be
\label{2055} {{{\partial}X^{\u m}}\over{{\partial}{\tau}}}\not= 0; \qquad
rank(D_{\hat q}{\Th}^{\u \mu})=n.  \ee The variation of \p{205} with
respect to $P_{{\u m}{\hat q}}$ yields the geometrodynamical constraint
\p{204} as a superfield equation. In the case of $N=1$ superparticles
\p{204} defines a non--minimal embedding of the superparticle worldline
into the target superspace, and \p{205} consistently describes the
dynamics of the $N=1$ superparticles in $D=3,4,6$ and $10$.

       The superfield action \p{205} is invariant with respect to local
worldline superdiffeomorphisms \be \label{206} \tau \rightarrow \tau
'(\tau ,\eta ),\qquad \eta _{\hat q}\rightarrow \eta '_{\hat q}(\tau ,\eta
), \ee which are restricted to transform the flat supervielbein form
$e_{\tau}$ \p{201} homogeneously:  \be \label{207} e'_{\tau}=W(\tau ,\eta
)e_{\tau}.  \ee This requirement imposes the constraint on the
superreparametrization functions $\tau '(\tau ,\eta )$ and $\eta '_{\hat
q}(\tau ,\eta )$:  \be \label{208} D_{   {\hat q}}\tau '-i(D_{   {\hat q}}
{\eta '}_{\hat p}){\eta '}_{\hat p}=0, \ee and ensures that the covariant
derivatives $D_{   {\hat q}}$ transform homogeneously as well \be
\label{209} D'_{   {\hat q}}=(D'_{   {\hat q}}{\eta}_{\hat p})D_{   {\hat
p}}.  \ee In the infinitesimal form these transformations are described by
a single superfield parameter $\L $ \cite{gs92,ghs}:  $$ \d \tau
={\L}-{1\over 2}{\eta}_{\hat q}D_{   {\hat q}} {\L};\qquad \d \eta _{\hat
q}=-{i\over 2}D_{   {\hat q}}{\L}; $$ \be \label{210} \d D_{   {\hat
q}}=-{1\over 2}{{\partial \L }\over{\partial \tau }} D_{   {\hat
q}}+{i\over 4}[D_{   {\hat q}},D_{{\hat p}}]{\L} D_{{\hat p}}; \ee $$ \ \d
(d\tau d^{D-2}\eta )=(1-{n\over 2}) {{\partial \L }\over{\partial \tau }}
(d\tau d^{D-2}\eta ).  $$ The invariance of the action \p{205} under
\p{210} requires the following transformation properties of the Lagrange
multiplier:  \be \label{211} \d P_{{\u m}{\hat
q}}={{n-1}\over{2}}{{\partial \L }\over{\partial \tau }} P_{{\u m}{\hat
q}}+{i\over 4}[D_{   {\hat q}},D_{   {\hat p}}]{\L} P_{{\u m}{\hat p}}.
\ee If the dimension of space--time is more than $3$, the action \p{205}
possesses \cite{gs92} an infinitely reducible symmetry \cite{bnsv} \be
\label{212} \d P_{{\u m}{\hat q}}=D_{   {\hat p}}({\bar \xi}_ {{\hat
p}{\hat q}{\hat s}}{\G}^{\u m}D_{   {\hat s}}{\Th}) \ee with the spinor
parameter ${\xi}^{     {\mu}}_{{\hat p}{\hat q}{\hat s}}$, which is
symmetric and traceless in ${\hat p}, {\hat q}, {\hat s}$. This symmetry
allows one to gauge away all $P_{{\u m}{\hat q}}$ components except $$
p_{\u m}={1\over{(D-2)!}} {\epsilon}_{{\hat q}_{_1},...,{\hat
q}_{_{D-2}}}D_{{\hat q}_{_1}} ...  D_{{\hat q}_{_{D-3}}}P_{{\u m}{\hat
q}_{_{D-2}}}\vert _{\eta =0}, $$ which satisfies the equation ${{\partial
p_{\u m}}\over{\partial \tau }}=0$ and plays the role of the superparticle
momentum. After elimination of all auxiliary fields the dynamical content
of the action \p{205} coincides with that of the
Casalbuoni--Brink--Schwarz superparticle in $D=3,4,6$ and $10$.

\section{$N=2$ superparticle in $D=3$}

        Doubly supersymmetric action for the $N=2~ D=3$ superparticle in
	terms of $n=2$ worldline superfields, constructed in \cite{gs2},
consists of the geometrodynamical term only:  \be \label{301} S_{GS}=\int
d\tau d^{2}\eta P_{{\hat {\u m}}{\hat q}} (D_{   {\hat q}}X^{\hat{\u m}}-
i(D_{   {\hat q}}{\Th}^{{\hat {\u \mu}}{\hat A}}) {\g}^{\hat {\u
m}}_{{\hat {\u \mu}}{\hat {\u \nu}}} {\Th}^{{\hat {\u \nu}}{\hat A}}).
\ee Here ${\hat {\u m}}=0,1,2$ is the $D=3$ vector index, ${\hat {\u
\mu}},{\hat {\u \nu}}=1,2$ are indices of three--dimensional Majorana
spinors, ${\g}^{\hat {\u m}}_{{\hat {\u \mu}}{\hat {\u \nu}}}$ are $D=3$
Dirac matrices in the real representation, ${\hat q}$ and ${\hat A}$ are
indices, corresponding to the local $n=2$ worldline supersymmetry of the
worldline and the global $N=2$ supersymmetry of the target space,
respectively.

         In this section we will show that the action \p{301} can be
	 obtained from the four--dimensional action \p{205} by dimensional
reduction and upon eliminating some of the auxiliary fields.

         We start with studying the structure of the geometrodynamical
	 constraint of the $N=2~ D=3$ superparticle and show that it does
not put the theory on the mass shell. In order to do this it is convenient
to rewrite this constraint in terms of  superinvariant one--forms:  \be
\label{302} \Pi ^{\hat {\u m}}\equiv dX^{\hat{\u m}}- id{\Th}^{{\hat {\u
\mu}}{\hat A}} {\g}^{\hat {\u m}}_{{\hat {\u \mu}}{\hat {\u \nu}}}
{\Th}^{{\hat {\u \nu}}{\hat A}}= e_{\tau}{\Pi}_{\tau}^{\hat {\u m}}, \ee
\be \label{303} \Pi ^{{\hat {\u \mu}}{\hat A}}=d\Th ^{{\hat {\u \mu}}{\hat
A}}\equiv e_{\tau}{\Pi}_{\tau}^{{\hat {\u \mu}}{\hat A}} +e_{\hat q}\Pi
^{{\hat {\u \mu}}{\hat A}}_{   {\hat q}}, \ee where $e_{\tau}$ and
$e_{\hat q}$ are intrinsic vielbeins defined in \p{201}, and $$
{\Pi}_{\tau}^{\hat{\u m}}={{\partial X^{\hat{\u m}}}\over{\partial \tau
}}- i{{\partial {\Th}^{{\hat {\u \mu}}{\hat A}}}\over{\partial \tau }}
{\g}^{\hat {\u m}}_{{\hat {\u \mu}}{\hat {\u \nu}}} {\Th}^{{\hat {\u
\nu}}{\hat A}}, $$ \be \label{304} {\Pi}_{\tau}^{{\hat {\u \mu}}{\hat A}}=
{{\partial {\Th}^{{\hat {\u \mu}}{\hat A}}}\over{\partial \tau }};~ \Pi _{
{\hat q}}^{{\hat {\u \mu}}{\hat A}}= D_{   {\hat q}}{\Th}^{{\hat {\u
\mu}}{\hat A}}.  \ee

     Selfconsistency conditions for the Eqs. \p{302} and \p{303} are:  \be
\label{305} -id{\Th}^{{\hat {\u \mu}}{\hat A}} {\g}^{\hat {\u m}}_{{\hat
{\u \mu}}{\hat {\u \nu}}} d{\Th}^{{\hat {\u \nu}}{\hat A}}=
d(e_{\tau}{\Pi}_{\tau}^{\hat {\u m}}), \ee \be \label{306}
d(e_{\tau}\Pi_{\tau}^{{\hat {\u \mu}}{\hat A}}) +d(e_{\hat q }\Pi ^{{\hat
{\u \mu}}{\hat A}}_{   {\hat q}})=0.  \ee Expanding the l. h. s. of
\p{305} in components \p{303} and using the ``constraints" on $e_{\tau}$
and $e_{\hat q}$ $$ de_{\tau}=-ie_{\hat q}e_{\hat q},~de_{\hat q}=0, $$
which follow from \p{201}, one can rewrite \p{305} and \p{306} as follows
\be \label{307} \Pi _{   \{ {\hat q}}^{{\hat {\u \mu}}{\hat A}} {\g}^{\hat
{\u m}}_{{\hat {\u \mu}}{\hat {\u \nu}}} \Pi _{   {\hat p}\} }^{{\hat {\u
\nu}}{\hat A}}={1\over 2} \d _{{\hat q}{\hat p}} \Pi _{{\hat s}}^{{\hat
{\u \mu}}{\hat A}} {\g}^{\hat {\u m}}_{{\hat {\u \mu}}{\hat {\u \nu}}} \Pi
_{{\hat s}}^{{\hat {\u \nu}}{\hat A}}, \ee \be \label{308}
\Pi_{\tau}^{\hat {\u m}}= \Pi _{    {\hat q}}^{{\hat {\u \mu}}{\hat A}}
{\g}^{\hat {\u m}}_{{\hat {\u \mu}}{\hat {\u \nu}}} \Pi _{   {\hat
q}}^{{\hat {\u \nu}}{\hat A}}.  \ee Eq. \p{308} is called {\it the twistor
constraint} since it expresses the vector $\Pi ^{\hat {\u m}}$ as a
bilinear combination of commuting spinors $\Pi _{    {\hat s}}^{{\hat {\u
\mu}}{\hat A}}$. Eqs. \p{307} are algebraic constraints, relating the
bosonic spinor superfields $\Pi _{    {\hat s}}^{{\hat {\u \mu}}{\hat A}}$
to each other.

      Assume that spinors $\Pi _{    {\hat s}}^{{\hat {\u \mu}}{\hat 1}}$
form a complete non--degenerate spinor basis in $D=3$ (i.e.  $det
{\Pi}_{\hat s}^{\hat{\u \mu}{\hat 1}}\not= 0$). Then $\Pi _{    {\hat
s}}^{{\hat {\u \mu}}{\hat 2}}$ can be represented as a linear combination
of basic spinors:  \be \label{309} \Pi _{    {\hat p}}^{{\hat {\u
\mu}}{\hat 2}}=a_{{\hat p}{\hat q}} \Pi _{    {\hat q}}^{{\hat {\u
\mu}}{\hat 1}}.  \ee Substituting this expression into \p{307} and taking
into account the linear independence of three non--vanishing vectors $\Pi
_{    {\hat p}}^{{\hat {\u \mu}}{\hat A}} {\g}^{\hat {\u m}}_{{\hat {\u
\mu}}{\hat {\u \nu}}} \Pi _{   {\hat q}}^{{\hat {\u \nu}}{\hat A}}$, one
obtains that the constraint \p{307} reduces to conditions on the
coefficients $a_{{\hat p}{\hat q}}$:  \be \label{310} a_{\{ {\hat p}{\hat
t}}a_{{\hat q}\} {\hat s}}+ \d _{\{ {\hat p}{\hat t}}\d _{{\hat q}\} {\hat
s}}={1\over 2} \d _{{\hat p}{\hat q}}(a_{{\hat t}{\hat u}}a_{{\hat s}
{\hat u}}+ \d _{{\hat t}{\hat s}}) \ee (on the l. h. s. of this expression
only the indices ${{\hat p},{\hat q}}$ are symmetrized). The general
solution to the system \p{310} has the following form:  \be \label{311}
a_{{\hat p}{\hat q}} =\pm \epsilon _{{\hat p}{\hat q}}.  \ee One can
select the sign (say, $+$) without the loss of generality, the system
\p{307} reducing to \be \label{312} \Pi _{    {\hat p}}^{{\hat {\u
\mu}}{\hat 2}}=\epsilon _{{\hat p}{\hat q}} \Pi _{    {\hat q}}^{{\hat {\u
\mu}}{\hat 1}}.  \ee Hitting \p{312} by the Grassmann covariant
derivatives $D_{   {\hat q}}$, one obtains the constraint which expresses
the field $\epsilon _{{\hat q}{\hat p}}D_{   {\hat q}}D_{   {\hat
p}}{\Th}^{{\hat {\u \mu}}{\hat A}}$ in terms of $\Pi_{\tau}^{{\hat {\u
\mu}}{\hat A}}$:  \be \label{313} \epsilon _{{\hat q}{\hat p}}D_{   {\hat
q}}D_{   {\hat p}}{\Th}^{\hat{\u \mu}{\hat A}}= 2i\epsilon ^{{\hat A}{\hat
B}}{\Pi}_{\tau}^{{\hat {\u \mu}}{\hat B}} \ee Taking into account \p{312}
one gets the twistor constraint in the following form:  \be \label{314}
{\Pi}_{\tau}^{\hat {\u m}}= 2\Pi _{    {\hat s}}^{{\hat {\u \mu}}{\hat 1}}
{\g}^{\hat {\u m}}_{{\hat {\u \mu}}{\hat {\u \nu}}} \Pi _{   {\hat
s}}^{{\hat {\u \nu}}{\hat 1}}; \ee One can see that the  geometrodynamical
constraint does not contain dynamical equations. It includes only
constraints which express higher components of the superfields $X^{\hat
{\u m}}$ and ${\Th}^{{\hat {\u \mu}}{\hat A}}$ in terms of their leading
components. This means that the superfield $P_{{\hat {\u m}}{\hat q}}$
does not contain any dynamical degrees of freedom except the superparticle
momentum \be \label{315} p^{{\hat {\u m}}}= {1\over 2}\epsilon _{{\hat
q}{\hat p}}D_{{\hat q}} P_{   {\hat p}}^{{\hat {\u m}}}\vert _{\eta =0}.
\ee All other $P_{{\hat p}}^{{\hat {\u m}}{   }}$ components are auxiliary
fields which can be eliminated either by explicit solution of constraints
or by fixing gauges of available local symmetries \cite{gs2}.

     In order to clarify the structure of the local symmetries and with
the purpose of the generalization to higher space--time dimensions we
shall obtain a superfield action for the $N=2 ~D=3$ superparticle once
again by the dimensional reduction of the doubly supersymmetric
formulation of an $N=1 ~D=4$ superparticle considered in the previous
Section. A reason for this is that the symmetry structure of the doubly
supersymmetric formulation of $N=1 ~D=4$ is known and it is retained after
dimensional reduction.

     The $D=4$ action \p{205}, written in terms of two--component Weyl
     spinors, is \be \label{316} S_{D=4}^{N=1}=\int d\tau d^{2}\eta
P_{{{\u m}}{\hat q}} (D_{   {\hat q}}X^{{\u m}}- i(D_{   {\hat
q}}{\Th}^{{{\u \a}}}) {\s}^{{\u m}}_{{{\u \a}}{\dot {\u \a}}} {\bar
\Th}^{{\dot {\u \a}}}- i(D_{   {\hat q}}{\bar \Th}_{{\dot{\u \a}}}) {\bar
\s}^{{\u m}{\dot{\u \a}}{{\u \a}}} {\Th}_{{{\u \a}}}).  \ee

Here ${\u m}=0,1,2,3$ is the $D=4$ vector index and ${\u \a} ,{\u{\dot
\a}}$ are indices of the fundamental representations of $SL(2,{\bf C})$.

      Performing the dimensional reduction of \p{316} to $D=3$, we require
the coordinate $X^{\u 2}$ to be constant and the corresponding component
of the particle momentum to be zero, i.e. (see \p{315}) ${\epsilon}_{{\hat
p}{\hat q}}D_{\hat p}P_{{\u 2}{\hat q}}\vert _{\eta =0}=0$.  Then,
redefining the spinors and the $\s $--matrices:  $$ 0,1,3 \rightarrow
{\hat 0}, {\hat 1}, {\hat 2} $$ $$ {\s}^{{\u 0}}_{{{\u \a}}{\dot {\u
\a}}}\rightarrow {\g}^{\hat {\u 0}}_{{\hat {\u \mu}}{\hat {\u \nu}}},\quad
{\s}^{{\u 1}}_{{{\u \a}}{\dot {\u \a}}}\rightarrow {\g}^{\hat {\u
1}}_{{\hat {\u \mu}}{\hat {\u \nu}}},\quad {\s}^{{\u 3}}_{{{\u \a}}{\dot
{\u \a}}}\rightarrow {\g}^{\hat {\u 2}}_{{\hat {\u \mu}}{\hat {\u
\nu}}},\quad {\s}^{{\u 2}}_{{{\u \a}}{\dot {\u \a}}}\rightarrow
i{\epsilon}_{{\hat {\u \mu}}{\hat {\u \nu}}}, $$ $$ {\Th}^{{\u
\a}}\rightarrow {1\over{\sqrt 2}} ({\Th}^{{\hat {\u \mu}}{\hat
1}}+i{\Th}^{{\hat {\u \mu}}{\hat 2}}),\qquad {\bar \Th}^{\dot{\u
\a}}\rightarrow {1\over{\sqrt 2}} ({\Th}^{{\hat {\u \mu}}{\hat
1}}-i{\Th}^{{\hat {\u \mu}}{\hat 2}}), $$ we obtain a $D=3$ functional \be
\label{317} S_{D=3}^{N=2}=\int d\tau d^{2}\eta [P_{{\hat {\u m}}{\hat q}}
(D_{   {\hat q}}X^{\hat{\u m}}- i(D_{   {\hat q}}{\Th}^{{\hat {\u
\mu}}{\hat A}}) {\g}^{\hat {\u m}}_{{\hat {\u \mu}}{\hat {\u \nu}}}
{\Th}^{{\hat {\u \nu}}{\hat A}})- iP_{{{\u 2}}{\hat q}} (D_{   {\hat
q}}{\Th}^{{\hat {\u \mu}}{\hat A}}) {\epsilon}^{{\hat A}{\hat B}}
{\Th}_{{\hat {\u \mu}}}^{{\hat B}})], \ee where one should take into
account that the Lagrange multiplier $P_{{{\u 2}}{\hat q}}$ is now a
restricted superfield. It must satisfy the equation $D_{   {\hat
q}}P_{{{\u 2}}{\hat q}}=0$, which follows from the action \p{316} as a
result of its variation with respect to $X^{\u 2}$. The general solution
to this equation is:  \be \label{318} P_{{{\u 2}}{\hat q}}=iD_{   {\hat
p}}Q_{{\hat p}{\hat q}}+ {1\over 2}{\epsilon}_{{\hat q}{\hat
p}}{\eta}_{\hat p} [{\epsilon}_{{\hat s}{\hat t}}D_{   {\hat s}}P_{{{\u
2}}{\hat t}}]\vert _{\eta =0}, \ee where $Q_{{\hat p}{\hat q}}$ is
symmetric and traceless in ${\hat p},~{\hat q}$.  The expression in
brackets is nothing but the second component of the $D=4$ particle
momentum $p_{\u 2}$ which we put to zero in the course of dimensional
reduction. Thus one can rewrite $P_{{{\u 2}}   {\hat q}}$ in terms of
unrestricted bosonic superfield $Q_{{\hat p}{\hat q}}$ only \be
\label{319} P_{{{\u 2}}{\hat q}}= iD_{   {\hat p}}Q_{{\hat p}{\hat q}}.
\ee Substituting this expression back into \p{317} and performing
integration by parts of the second term, one gets \be \label{320}
S_{D=3}^{N=2}=\int d\tau d^{2}\eta [P_{{\hat {\u m}}{\hat q}} (D_{   {\hat
q}}X^{\hat{\u m}}- i(D_{   {\hat q}}{\Th}^{{\hat {\u \mu}}{\hat A}})
{\g}^{\hat {\u m}}_{{\hat {\u \mu}}{\hat {\u \nu}}} {\Th}^{{\hat {\u
\nu}}{\hat A}})+ \ee $$ Q_{{\hat p}{\hat q}} (D_{   {\hat p}}{\Th}^{{\hat
{\u \mu}}{\hat A}}) {\epsilon}^{{\hat A}{\hat B}} (D_{   {\hat
q}}{\Th}_{{\hat {\u \mu}}}^{{\hat B}})].  $$ One can see that \p{317}
differs from the action \p{301} by the second term.  The action \p{320}
possesses the $n=2$ local worldline supersymmetry and is also invariant
under local transformations of the Lagrange multipliers $P_{{\hat {\u
m}}{\hat q}}$ and $Q_{{\hat p}{\hat q}}$:  \be \label{321} \d P_{{\u
m}{\hat q}}=D_{   {\hat p}} ({\xi}^{     {\hat {\u \mu}}{\hat A}}_{{\hat
p}{\hat q}{\hat s}} {\g}^{\u m}_{\hat{{\u \mu}}{\hat{\u \nu}}} D_{   {\hat
s}}{\Th}^{{{\hat{\u \nu}}{\hat A}}}),~~ \d Q_{{\hat p}{\hat q}}=-{i\over
2}({\xi}^{     {\hat {\u \mu}}{\hat A}}_ {{\hat p}{\hat q}{\hat
s}}{\epsilon}^{\hat{{A}}{\hat{B}}} D_{   {\hat s}}{\Th}_{\hat{\u
\mu}}^{\hat B}), \ee \be \label{322} \d Q_{{\hat p}{\hat q}}= D_{   {\hat
s}}({\Xi}_{{\hat p}{\hat q}{\hat s}}), \ee where the worldline superfield
parameters ${\xi}^{     {\hat {\u \mu}}{\hat A}}_{{\hat p}{\hat q}{\hat
s}}$ and ${\Xi}_{{\hat p}{\hat q}{\hat s}}$ are completely symmetric and
traceless in ${\hat p},~{\hat q}~,{\hat s}$.

       These symmetries allow one to eliminate all the components of
$Q_{{\hat p}{\hat q}}$ and $P_{{\u m}{\hat q}}$ except \p{315}.  The
remaining action is equivalent to the Casalbuoni--Brink--Schwarz
superparticle action. Additional spinor constraints obtained by varying
\p{320} with respect to $Q_{{\hat p}{\hat q}}$ are consequences of the
geometrodynamical condition. Thus the second term in \p{320} can be
dropped away by use of the symmetry \p{322} and the part of \p{321}
without the loss of any dynamical information about the superparticle
motion and one gets the action \p{301}. Note, however, that in contrast to
\p{301} where a local symmetry structure is hidden at the component level,
the action \p{320} possesses all gauge symmetries in an explicit
superfield form, which will be helpful for the construction and the
analysis of $N=2~ D=4$ and $D=6$ superparticle actions.

\section{$N=2 ~D=4$ superparticle action}

      In this Section the $N=2, ~D=4$ superparticle action will be
      obtained by the dimensional reduction of the action \p{205} in
$D=6$, which now takes the form \be \label{401} S_{D=6}^{N=1}=\int d\tau
d^{4}\eta P_{{{\u M}}{\hat q}} (D_{   {\hat q}}X^{{\u M}}- i(D_{   {\hat
q}}{\Th}^{{{\u \mu}}{\dot I}}) {\g}^{{\u M}}_{{\u \mu}{\dot {\u \nu}}}
{\bar \Th}^{{\dot {\u \nu}}}_{\dot I}).  \ee

       Here ${\u M}$ is the $SO(1,5)$ vector index, ${\hat q}$ is the
       index of the $n=4$ worldline {\susy } algebra.  The
$SU(2)$--Majorana spinor Grassmann coordinates ${\Th}^{{{\u \mu}}{\dot
I}}$ carry the $SU^{*}(4)$ index ${\u \mu}$ as well as the $SU(2)$--index
${\dot I}$ and satisfy the condition \cite{kgtown}:  $$ {\bar {{\Th}^{{{\u
\mu}}}_{\dot I}}}\equiv {\bar \Th}^{{\dot {\u \mu}}{\dot I}}= C^{\dot{\u
\mu}}_{~{\u \nu}} {\varepsilon}^{{\dot I}{\dot J}}{\Th}_{\dot J}^{\u \nu},
$$ where $C^{\dot{\u \mu}}_{~{\u \nu}}$ is a $D=6$ charge conjugation
matrix $$ C^{\dot{\u \mu}}_{~{\u \nu}}=\left( \begin{array}{cccc} 0 & 1 &
0 & 0 \\ -1& 0 & 0 & 0 \\ 0 & 0 & 0 & 1 \\ 0 & 0 & -1 & 0 \\ \end{array}
\right) $$ (index ${\dot{\u \mu}}$ corresponds to a conjugate $SU^{*}(4)$
representation).  ${\g}^{{\u M}}_{{\u \mu}{\dot {\u \nu}}}$ are the $D=6$
``Pauli" matrices, for which we choose the following realization \be
\label{402} {\g}^{\u m}=\left( \begin{array}{cc} 0 & {\s}^{\u m}_{{\u
\a}{\dot{\u \a}}} \\ {\bar \s}^{{\u m}{{\dot{\u \a}}{{\u \a}}}} & 0 \\
\end{array} \right), \quad {\g}^{\u 4}=\left( \begin{array}{cc}
-i{\d}^{{\u \b}}_{{\u \a}} & 0 \\ 0 & -i{\d}^{\dot{\u \a}}_{\dot{\u \b}}
\\ \end{array} \right), \quad {\g}^{\u 5}=\left( \begin{array}{cc}
-{\d}^{{\u \b}}_{{\u \a}} & 0 \\ 0 & {\d}^{\dot{\u \a}}_{\dot{\u \b}} \\
\end{array} \right), \ee where ${\u m}=0,1,2,3$ is the $SO(1,3)$ vector
index; ${\u \a}$, ${\u{\dot \a}}$ are $SL(2,C)$ indices and  $\sigma ^{\u
m}_{{\u \a}{\u{\dot \a}}}$ are the Pauli matrices. Decomposing the
$SU(2)$--Majorana spinors in the action \p{401} into a pair of
two--component $D=4$ Weyl spinors $$ {\Th}^{{\u \mu}{\dot I}}=({\Th}^{{\u
\a}{\dot I}},{\bar \Th}^{\dot I}_{\dot{\u \a}}), $$ substituting the
realization \p{402} and putting $~X^{{\u 4},{\u 5}}=const$, $~p^{{\u
4},{\u 5}}=0$, one obtains the following action reduced to $D=4$ $$
S_{D=4}^{N=2}=\int d\tau d^{4}\eta [P_{{{\u m}}{\hat q}} (D_{   {\hat
q}}X^{{\u m}}- i(D_{   {\hat q}}{\Th}^{{{\u \a}{\dot I}}}) {\s}^{{\u
m}}_{{{\u \a}}{\dot {\u \b}}} {\bar \Th}^{{\dot {\u \b}}}_{\dot I}- i(D_{
{\hat q}}{\bar \Th}_{{\dot{\u \a}}{\dot I}}) {\bar \s}^{{\u m}{\dot{\u
\a}}{{\u \b}}} {\Th}^{\dot I}_{{{\u \b}}})- $$ \be \label{4035}
i{\pi}_{{\hat q}} (D_{   {\hat q}}{\Th}^{{{\u \a}{\dot I}}}){\Th}_{{{\u
\a}}{\dot I}}- i{\bar \pi}_{{\hat q}} (D_{   {\hat q}}{\bar \Th}^{{\dot{\u
\a}}}_{\dot I}){\bar \Th}_{{\dot {\u \a}}}^{\dot I}].  \ee In \p{4035} we
introduced the complex superfield ${\pi}_{\hat q}{\equiv}P_{{\u 5}{\hat
q}}+iP_{{\u 4}{\hat q}}$, which satisfies the equation $D_{\hat
q}{\pi}_{\hat q}=0$ following from the variation of the \p{401} with
respect to $X^{\u 4}$ and $X^{\u 5}$. The general solution to this
equation has the form $$ {\pi}_{{\hat q}}=iD_{{\hat p}}Q_{{\hat p}{\hat
q}}+{\epsilon}_{{\hat q}{\hat p}{\hat s}{\hat t}} {\eta}_{\hat
p}{\eta}_{\hat s}{\eta}_{\hat t}(p_{\u 5}+ip_{\u 4}), $$ where $Q_{{\hat
p}{\hat q}}$ is an unrestricted superfield, which is symmetric and
traceless in ${\hat p}$ and ${\hat q}$; $p_{\u 4}$ and $p_{\u 5}$ are
components of the $D=6$ superparticle momentum $$ p_{\u M}={1\over
24}{\epsilon}_{{\hat p}{\hat q}{\hat s}{\hat t}} D_{{\hat p}}D_{{\hat
q}}D_{{\hat s}}P_{{{\u M}}{\hat t}} \vert _{\eta =0}, $$ which we put to
zero. So, ${\pi}_{{\hat q}}=iD_{   {\hat p}}Q_{{\hat p}{\hat q}}$.

      Inserting this expression into the action \p{4035} and performing
      integration by parts of the last two terms, one gets the action in
      the form similar to the action of the $N=2 ~D=3$ superparticle
considered in the previous Section:  \be \label{404} S_{D=4}^{N=2}=\int
d\tau d^{4}\eta  [P_{{{\u m}}{\hat q}} (D_{   {\hat q}}X^{{\u m}}- i(D_{
{\hat q}}{\Th}^{{{\u \a}{\dot I}}}) {\s}^{{\u m}}_{{\u \a}{\dot {\u \b}}}
{\bar \Th}^{{\dot {\u \b}}}_{\dot I}- i(D_{   {\hat q}}{\bar
\Th}_{{\dot{\u \a}}{\dot I}}) {\bar \s}^{{\u m}{\dot{\u \a}}{\u \b}}
{\Th}^{\dot I}_{{{\u \b}}}+ \ee $$ {Q}_{{\hat q}{\hat p}} (D_{   {\hat
q}}{\Th}^{{{\u \a}{\dot I}}}) (D_{   {\hat p}}{\Th}_{{{\u \a}{\dot I}}})+
c. c.] $$

      The action obtained  possesses manifest $n=4$ local worldline
supersymmetry \p{210} and is invariant under the following transformations
of the Lagrange multipliers:  $$ \d P_{{\u m}{\hat q}}= {1\over 2}D_{
{\hat p}} ({\xi}^{     {{\u \a}}{\dot I}}_{{\hat p}{\hat q}{\hat s}}
{\s}^{\u m}_{{{\u \a}}{\dot{\u \b}}} D_{   {\hat s}}{\bar \Th}^{{\dot {\u
\b}}}_{{\dot I}})+ {1\over 2}D_{   {\hat p}} ({\bar \xi}_{{{\u \a}}{\dot
I}{{\hat p}{\hat q}{\hat s}}} {\bar \s}^{{\u m}{{\dot{\u \a}}{{\u \b}}}}
D_{   {\hat s}}{\Th}_{{{\u \b}}}^{{\dot I}}), $$ \be \label{405} \d
Q_{{\hat p}{\hat q}}=-i({\bar \xi}_{{\hat {\u \a}}{\dot I} {{\hat p}{\hat
q}{\hat s}}} D_{   {\hat s}}{\Th}^{{\dot{\u \a}}{\dot I}}), \ee \be
\label{406} \d Q_{{\hat p}{\hat q}}= D_{   {\hat s}}({\Xi}_{{\hat p}{\hat
q}{\hat s}}), \ee with the parameters ${\Xi}_{{\hat p}{\hat q}{\hat s}}$
and ${\xi}^{     {{\u \a}}{\dot I}}_{{\hat p}{\hat q}{\hat s}}$ completely
symmetric and traceless in ${\hat p},{\hat q}$ and ${\hat s}$.  The
      variation of \p{404} with respect to the Lagrange multiplier $P_{{\u
m}{\hat q}}$ gives the  geometrodynamical constraint  of the $N=2,~D=4$
superparticle, which can be written in the form \be \label{407} \Pi ^{\u
m}\equiv dX^{{\u m}}-id{\Th}^{{{\u \a}{\dot I}}} {\s}^{{\u m}}_{{{\u
\a}}{\dot {\u \b}}} {\bar \Th}^{{\dot {\u \b}}}_{\dot I}- id{\bar
\Th}_{{{\dot{\u \a}}}{\dot I}} {\bar \s}^{{\u m}{\dot{\u \a}}{{\u \b}}}
{\Th}^{\dot I}_{\u \b}=e_{\tau}{\Pi}_{\tau}^{\u m}, \ee \be \label{408}
\Pi^{{\u \a}{\dot I}}\equiv d{\Th}^{{{\u \a}{\dot
I}}}=e_{\tau}{\Pi}_{\tau}^{{\u \a}{\dot I}}+ e_{\hat q}{\Pi}_{{\hat
q}}^{{\u \a}{\dot I}}.  \ee

     For the analysis of this constraint it is convenient to use the
     complex $SU(2)$ vector parametrization for $n=4$ worldline Grassmann
coordinates:  $$\eta _{\hat q}\rightarrow (\eta _i; \bar{(\eta
_{i})}={\bar \eta}^{   i}).$$ The flat worldline superinvariant one--forms
\p{201} in this notation are \be \label{409} e_{\tau}=d{\tau}-{i\over
2}d{\eta}^{   i}{\bar \eta}_i- {i\over 2}d{\bar \eta}_i{\eta}^{   i},\quad
e^{   i}=d{\eta}^{   i},\quad {\bar e}_i=d{\bar \eta}_i, \ee so one can
rewrite \p{408} as:  \be \label{410} \Pi^{{\u \a}{\dot
I}}=e_{\tau}{\Pi}_{\tau}^{{\u \a}{\dot I}}+ e^{   i}{\Pi}_{   i}^{{\u
\a}{\dot I}}+ {\bar e}_i{\tilde \Pi}^{{\u \a}{\dot I}i}; \ee $$ {\Pi}_{
i}^{{\u \a}{\dot I}}=D_{   i}{\Th}^{{\u \a}{\dot I}};\quad {\tilde
\Pi}^{{\u \a}{\dot I}i}={\bar D}^i{\Th}^{{\u \a}{\dot I}}; $$ where  $D_{
i}={{\partial}\over{\partial \eta ^{   i}}}+ i{\bar
\eta}_{i}{{\partial}\over{\partial \tau }}$ and ${\bar
D}^i=-{{\partial}\over{\partial {\bar \eta}_i}}- i{\eta}^{
i}{{\partial}\over{\partial \tau }}$ are complex conjugate Grassmann
covariant derivatives, which satisfy the following (anti)commutation
relations \be \label{411} \{ D_{   i}, {\bar D}^j\} =-2i{\d}^j_i
{{\partial}\over{\partial \tau }},\qquad \{ D_{   i},D_{   j}\} =0,\qquad
[D_{   i},{{\partial}\over{\partial \tau }}] =0.  \ee

          Selfconsistency conditions for \p{407} and \p{410} $$
-2id{\Th}^{{\u \a}{\dot I}}{\s}^{\u m}_{{\u \a}{\dot{\u \a}}}
d{\bar{\Th}}^{\dot{\u \a}}_{\dot I}= e_{\tau}d{\Pi}_{\tau}^{\u m}-ie^{
i}{\bar e}_j{\d}^j_i{\Pi}_{\tau}^{\u m}, $$ $$ e_{\tau}d{\Pi}_{\tau}^{{\u
\a}{\dot I}}-ie^{   i}{\bar e}_j{\d}^j_i {\Pi}_{\tau}^{{\u \a}{\dot
I}}+e^{   i}d{\Pi}^{{\u \a}{\dot I}}_{   i}+ {\bar e}_id{\tilde \Pi}^{{\u
\a}{\dot I}i}=0; $$ give algebraic constraints on the fields ${\Pi}^{{\u
\a}{\dot I}}_i$ and ${\tilde \Pi}^{{\u \a}{\dot I}i}$:  \be \label{412}
{\Pi}^{{\u \a}{\dot I}}_{   \{ i}{\s}^{\u m}_{{\u \a}{\dot{\u \a}}}
{\bar{\tilde \Pi}}^{\dot{\u \a}}_{   j\} {\dot I}}=0; \ee $$ {\Pi}^{{\u
\a}{\dot I}}_{   i}{\s}^{\u m}_{{\u \a}{\dot{\u \a}}}{\bar \Pi}_{
I}^{{\dot{\u \a}}j}+ {\tilde \Pi}^{{\u \a}j{\dot I}}{\s}^{\u m}_{{\u
\a}{\dot{\u \a}}} {\bar{\tilde \Pi}}^{\dot{\u \a}}_{   i{\dot I}}= $$ \be
\label{413} {1\over 2}{\d}^j_i ({\Pi}^{{\u \a}{\dot I}}_{   k} {\s}^{\u
m}_{{\u \a}{\dot{\u \a}}} {\bar \Pi}_{   {\dot I}}^{{\dot{\u \a}}k}+
{\tilde \Pi}^{{\u \a}k{\dot I}}{\s}^{\u m}_{{\u \a}{\dot{\u \a}}}
{\bar{\tilde \Pi}}^{\dot{\u \a}}_{   k{\dot I}}) \ee as well as the
twistor constraint \be \label{414} {\Pi}^{\u m}= {\Pi}^{{\u \a}{\dot I}}_{
k}{\s}^{\u m}_{{\u \a}{\dot{\u \a}}}{\bar \Pi}_{   {\dot I}}^{{\dot{\u
\a}}k}+ {\tilde \Pi}^{{\u \a}k{\dot I}}{\s}^{\u m}_{{\u \a}{\dot{\u \a}}}
{\bar{\tilde \Pi}}^{\dot{\u \a}}_{   k{\dot I}}.  \ee To solve the system
of the algebraic constraints \p{412} and  \p{413} we consider the spinors
${\Pi}^{{\u \a}{\dot 1}}_{   i}$ as a $D=4$ spinor basis and expand all
other spinor superfields in it:  \be \label{415} {\Pi}^{{\u \a}{\dot
2}}_i=a_i^{~j}{\Pi}_j^{{\u \a}{\dot 1}},\qquad {\tilde \Pi}^{{\u \a}i{\dot
I}}=b^{{\dot I}ij}{\Pi}_j^{{\u \a}{\dot 1}}, \ee $a_i^{~j}$ and $b^{{\dot
I}ij}$ being superfield coefficients.  Substituting  \p{415} into \p{412}
and \p{413} and assuming that the vectors ${\Pi}_i^{{\u \a}{\dot
1}}{\s}^{\u m}_{{\u \a}{\dot{\u \a}}} {\bar \Pi}_{j{\dot 1}}^{{\dot{\u
\a}}}$ are non--vanishing and linearly independent, one obtains that the
constraints reduce to the following restrictions on the coefficients $$
a_i^{~j}=a{\d}_i^j, \quad b^{1ij}=-{\bar a}e^{i\varphi
}{\epsilon}^{ij},\quad b^{2ij}=e^{i\varphi }{\epsilon}^{ij}, $$ where
$a\in {\bf C}$ and $\varphi \in {\bf R}$ are arbitrary superfields which
can be gauged away by the use of local $SU(2)$ transformations generated
by the fields $D_iD^i\L \vert _{\eta =0},~{1\over 2}[D_i,{\bar D}^i]\L
\vert _{\eta =0}$, being part of the local worldline superdiffeomorphisms.
So, one can rewrite the constraints for bosonic superfields \p{415} as
follows \be \label{416} {\Pi}^{{\u \a}{\dot 2}}_i=a{\Pi}_i^{{\u \a}{\dot
1}},\qquad {\tilde \Pi}^{{\u \a}i{\dot 1}}=-{\bar a}e^{i\varphi
}{\Pi}^{{\u \a}i{\dot 1}},\qquad {\tilde \Pi}^{{\u \a}i{\dot
2}}=e^{i\varphi }{\Pi}^{{\u \a}i{\dot 1}}.  \ee

	Hitting these constraints by fermionic covariant derivatives and
	using the algebra \p{411}, we obtain constraints on higher
components of the superfields $a$ and $\varphi $ and expressions for the
second order components of the superfield $\Th $ in terms of the leading
ones:  $$ D_{   i}a=0;~D_{   i}\varphi ={i\over{1+a{\bar a}}}(aD_{
i}{\bar a}+e^{-i\varphi }{\bar D}_{   i}a), $$ \be \label{4166} D_{
i}D^i{\Th}^{{\u \a}{\dot 1}}=-{{4ie^{-i\varphi }}\over{1+a{\bar a}}}
(a{{\partial \Th ^{{\u \a}{\dot 1}}}\over{\partial \tau }}- {{\partial \Th
^{{\u \a}{\dot 2}}}\over{\partial \tau }})+ i(D^i\varphi )D_{
i}{\Th}^{{\u \a}{\dot 1}}+ \ee $$ {i\over{1+a{\bar a}}}(aD^i{\bar
a}+e^{-i\varphi }{\bar D}^ia)D_{   i}{\Th}^{{\u \a}{\dot 1}}, $$ $$ {\bar
D}^iD^j{\Th}^{{\u \a}{\dot 1}}={1\over 2}{\epsilon}^{ij} [{{4ie^{-i\varphi
}}\over{1+a{\bar a}}} ({{\partial \Th ^{{\u \a}{\dot 1}}}\over{\partial
\tau }}+ {\bar a}{{\partial \Th ^{{\u \a}{\dot 2}}}\over{\partial \tau
}})+ {i\over{1+a{\bar a}}}({\bar a}{\bar D}^k{a}+e^{i\varphi }{D}^k{\bar
a})D_{   k}{\Th}^{{\u \a}{\dot 1}}]- $$ $$ {\s}^{{\hat a}ij} [{i\over
2}({\bar D}^k\varphi ){\s}^{{\hat a}~l}_kD_{   l}{\Th}^{{\u \a}{\dot 1}}],
$$ (${\hat a}$ is the $SO(3)$ index). We observe that no dynamical
equations appear in \p{416} and \p{4166}.  The twistor constraint \p{414}
acquires the form $$ {\Pi}^{\u m}= -(1+a{\bar a}){\Pi}^{{\u \a}{\hat 1}}_{
k}{\s}^{\u m}_{{\u \a}{\dot{\u \a}}}{\bar \Pi}_{   {\hat 1}}^{{\dot{\u
\a}}k}.  $$

      One can show that the additional constraints on the spinors
${\Pi}^{{\u \a}{\dot I}}_{   i}$ and ${\tilde \Pi}^{{\u \a}i{\dot 1}}$
obtained by the variation of \p{404}  with respect to $Q_{{\hat p}{\hat
q}}$ are identically satisfied if one takes into account \p{416}. The
Lagrange multiplier $Q_{{\hat p}{\hat q}}$  is a purely auxiliary degree
of freedom which can be completely gauged away by use of the
transformations \p{405} and \p{406}.  Their role is to ensure the
invariance of the action under the superfield transformations \p{405} and
\p{406} which simplifies the analysis of the physical degrees of freedom
of the model.  Using the symmetry \p{405} and the constraints contained in
the superfield equations for $X^{\u m}$ and ${\Th}^{{\u \a}{\dot I}}$, one
can reduce the superfield $P_{{\u m}   {\hat q}}$ to the form $$ {P}_{{\u
m}{\hat q}}= {\epsilon}_{{\hat q}{\hat p}{\hat s}{\hat t}} {\eta}_{\hat
p}{\eta}_{\hat s}{\eta}_{\hat t}p_{\u m}, $$ where $p_{\u m}$ is the
superparticle momentum, satisfying the equation ${{\partial p_{\u
m}}\over{\partial \tau }}=0$. The set of equations, remaining after
elimination of all auxiliary fields and an explicit solution of
constraints describe the dynamics of the $N=2 ~D=4$ Brink--Schwarz
superparticle.

\section{Doubly supersymmetric action of the $D=6$ Type $IIA$
superparticle.}

      When considering the $N=2,~D=6$ superparticle, one should
distinguish two different kinds of the target superspace, corresponding to
so--called $IIA$ and $IIB$ theories. In the former two fermionic spinor
coordinates have opposite chiralities, while in the latter they have the
same chirality.  The superfield action of the type discussed herein can be
constructed for the Type $IIA$ superparticle only. The reason is that  the
 geometrodynamical constraint does not put the $IIA$ theory on the mass
 shell, while, as we will show, the  geometrodynamical constraint of the
$IIB~ D=6$ superparticle produces dynamical equations.

We start with demonstration that the dimensional reduction of the doubly
supersymmetric action of $N=1~D=10$ superparticle to $D=6$ yields the
action of the Type $IIA$ theory.

      The superfield action \p{205} for $N=1$ superparticle in $D=10$ has
the following form \cite{gs92} \be \label{501} S^{N=1}_{D=10}=\int d\tau
d^{8}\eta P_{{\hat {\u M}}{\hat q}} (D_{   {\hat q}}X^{\hat{\u M}}- i(D_{
{\hat q}}{\Th}^{{\hat {\u \a}}}) {\g}^{\hat {\u M}}_{{\hat {\u \a}}{\hat
{\u \b}}} {\Th}^{{\hat {\u \b}}}), \ee where ${\hat {\u M}}=0,...,9$ is a
vector $SO(1,9)$ index, ${\hat {\u \a}},{\hat{\u \b}}=1,...,16$ are
indices of 16--component Majorana--Weyl spinors, ${\g}^{\hat {\u
M}}_{{\hat {\u \a}}{\hat {\u \b}}}$ are $16\times 16$ ten--dimensional
${\g}$--matrices, satisfying $$ {\g}^{\{ {\hat{\u M}}}_{{\u \a}{\u \b}}
{\tilde \g}^{{\hat{\u N}}\} {\u \b}{\u \g}}={\eta}^{{\hat{\u M}}{\hat{\u
N}}} {\d}^{\u \g}_{\u \a}, $$ ${\hat q}$ is the index, corresponding to
 the $n=8$ worldline supersymmetry. To reduce \p{501} to $D=6$ we
decompose the Majorana--Weyl spinor ${\Th}^{\hat{\u \a}}$ into two $D=6~
SU(2)$--Majorana spinors of opposite chiralities:  \be \label{502}
{\Th}^{\hat{\u \a}}=({\Th}^{(1){\u \mu}{\dot I}}, {\Th}^{(2)I}_{\u \mu}),
\ee ($D=6$ notation has been presented in the previous Section) and use
the following realization for $D=10$ matrices ${\g}^{\hat {\u M}}_{{\hat
{\u \a}}{\hat {\u \b}}}$:  ${\hat{\u M}}=({\u M},{\hat i})$, ${\u
M}=0,...,5$ is the $SO(1,5)$ index; ${\hat i}=6,7,8,9$ is an index of the
vector representation of $SO(4)$, corresponding to four compactified
space--time dimensions:  $$ {\g}^{\u M}_{{\hat{\u \a}}{\hat{\u
\b}}}=\left( \begin{array}{cc} {\epsilon}_{{\dot I}{\dot J}}{\g}^{\u
M}_{{{\u \mu}}{{\u \nu}}} & 0 \\ 0 & {\epsilon}_{{I}{J}}{\tilde \g}^{{\u
M}{{{\u \mu}}{{\u \nu}}}} \\ \end{array} \right); \quad {\tilde \g}^{{\u
M}{{\hat{\u \a}}{\hat{\u \b}}}}=\left( \begin{array}{cc}
-{\epsilon}^{{\dot I}{\dot J}}{\tilde \g}^{{\u M}{{{\u \mu}}{{\u \nu}}}} &
0 \\ 0 & {\epsilon}^{{I}{J}}{\g}^{\u M}_{{{\u \mu}}{{\u \nu}}} \\
\end{array} \right); \quad $$ \be \label{503} {\g}^{i}=\left(
\begin{array}{cc} 0 & {\bar \s}^{i}_{{\dot I}I}{\d}^{\u \mu}_{\u \nu} \\
{\s}^i_{I{\dot I}}{\d}^{\u \nu}_{\u \mu} & 0 \\ \end{array} \right); \quad
{\bar \g}^{i}=\left( \begin{array}{cc} 0 & -{\bar \s}^{{i}{{\dot
I}I}}{\d}^{\u \mu}_{\u \nu} \\ -{\s}^{iI{\dot I}}{\d}^{\u \nu}_{\u \mu} &
0 \\ \end{array} \right); \quad \ee where the matrices ${\s}^{i}_{I{\dot
I}}$ constitute a non--degenerate basis in the space of $2\times 2$
matrices, transformed under the representation of $SO(4)\sim SU(2)\times
{SU(2)}$ ($I,J$ are indices of $SU(2)$, ${\dot I},{\dot J}$ are indices of
another ${SU(2)}$ and $i$ is an index of the vector $SO(4)$
representation).  They satisfy the following relations:  $$ {\s}^{\{
i}_{I{\dot I}}{\bar \s}^{{j}\} {\dot I}J}= {\d}^{{i}{j}}{\d}^{J}_{I},\quad
{\bar \s}^{{\dot I}J\{ i}{\s}^{{j}\} }_{J{\dot J}}=
{\d}^{{i}{j}}{\d}^{\dot I}_{\dot J},\quad {\bar \s}^i_{{\dot
I}I}={\s}^i_{I{\dot I}}.  $$

   Substituting the representation \p{503} and performing the dimensional
   reduction to $D=6$, we obtain the following action in $D=6$ $IIA$
superspace:  \be \label{505} S_{D=6}^{N=2}=\int d\tau d^{8}\eta  [P_{{{\u
M}}{\hat q}} (D_{   {\hat q}}X^{{\u M}}- i(D_{   {\hat q}}{\Th}^{(1){{\u
\mu}{\dot I}}}) {\g}^{{\u M}}_{{{\u \mu}}{{\u \nu}}} {\Th}^{(1){\u
\nu}}_{\dot I}- i(D_{   {\hat q}} {\Th}_{{\dot{\u \mu}}}^{(2){I}}) {\tilde
\g}^{{\u M}{{\u \mu}{\u \nu}}} {\Th}^{(2)}_{{I}{\u \nu}})+ \ee $$ {Q}^{i
}_{{\hat q}{\hat p}} (D_{   {\hat q}}{\Th}^{(2)I}_{\u \mu}){\s}^i_{I{\dot
I}} (D_{   {\hat p}}{\Th}^{(1){\u \mu}{\dot I}})], $$ where ${Q}^{i
}_{{\hat q}{\hat p}}$ is a superfield Lagrange multiplier, which is
symmetric and traceless in ${\hat p}$ and ${\hat q}$.

       In addition to the local $n=8$ worldline supersymmetry the action
       \p{505} is invariant under the following transformations of the
Lagrange multipliers $P_{{\u M}{\hat q}}$ and ${Q}^{i    }_{{\hat q}{\hat
p}}$:  $$ \d P_{{\u M}{\hat q}}= D_{   {\hat p}} ({\xi}^{     {(1){\u
\mu}}{\dot I}}_{{\hat p}{\hat q}{\hat s}} {\g}^{\u M}_{{{\u \mu}}{{\u
\nu}}} D_{   {\hat s}}{\Th}^{(1){{\u \nu}}}_{{\dot I}})+ D_{   {\hat p}}
({\xi}^{     (2)I}_{{\u \mu}{\hat p}{\hat q}{\hat s}} {\tilde \g}^{{\u
M}{\u \mu}{\u \nu}} D_{   {\hat s}}{\Th}_{{{\u \nu}}I}^{(2)}); $$

\be \label{5065} \d Q^{    i}_{{\hat p}{\hat q}}= -i[({\xi}^{     (1){\dot
I}{\u \mu}} _{{\hat p}{\hat q}{\hat s}}{\bar \s}^i_{{\dot I}I} D_{   {\hat
s}}{\Th}_{{{\u \mu}}}^{(2)I})+ ({\xi}^{     (2){I}}_{{\u \mu} {{\hat
p}{\hat q}{\hat s}}}{\s}^i_{I{\dot I}} D_{   {\hat s}}{\Th}^{{{{\u
\mu}}}(1){\dot I}})]; \ee \be \label{506} \d Q^{    i}_{{\hat p}{\hat q}}=
D_{   {\hat s}}({\Xi}^{     i}_{{\hat p}{\hat q}{\hat s}}); \ee with the
parameters ${\Xi}_{{\hat p}{\hat q}{\hat s}}$, ${\xi}^{(1){{\u \mu}}{\dot
I}}_{{\hat p}{\hat q}{\hat s}}$.  and ${\xi}^{(2){I}}_{{\hat p}{\hat
q}{\hat s}{\u \mu}}$, which are completely symmetric and traceless with
respect to the $SO(8)$ indices.

       Variation of the action \p{505} with respect to the Lagrange
       multiplier $P_{{\u M}{\hat q}}$ gives the  geometrodynamical
constraint  of the $IIA$ $D=6$ superparticle, which we rewrite in terms of
superinvariant one--forms $$ {\Pi}^{\u M}_{IIA}\equiv dX^{{\u M}}-
id{\Th}^{(1){{\u \mu}{\dot I}}}{\g}^{{\u M}}_{{{\u \mu}}{{\u \nu}}}
{\Th}^{(1){\u \nu}}_{\dot I}- id{\Th}_{{\dot{\u \mu}}}^{(2){I}}{\tilde
\g}^{{\u M}{{\u \mu}{\u \nu}}} {\Th}^{(2)}_{{I}{\u \nu}}=e_{\tau}{\Pi}^{\u
M}_{{\tau}IIA}; $$ Now we demonstrate that this constraint does not
produce dynamical equations.  In order to analyze its structure we choose
the light--cone basis:  $$ X^{\u M}=(X^{++}=X^{\u 0}+X^{\u
5};~X^{--}=X^{\u 0}-X^{\u 5};~X^i);~i=1,2,3,4; $$ and decompose the
$SU^{*}(4)$ spinor index $\mu $ of the Grassmann target space coordinates
into a pair of  $SU(2)\times SU(2)$ indices $A$ and ${\dot A}$,
introducing the sign ``indices" $+,-$ (weights) of the $SO(1,1)$ subgroup
of $SU^{*}(4)$ \be \label{5062} {\Th}^{(1){\u \mu}{\dot I}}=
({\Th}^{(1)+A{\dot I}},{\Th}^{(1)-{\dot A}{\dot I}}),\qquad
{\Th}^{(2)I}_{\u \mu}=({\Th}^{(2)-I}_{~~A},{\Th}^{(2)+I}_{~~{\dot A}}).
\ee The  geometrodynamical constraint of $IIA$ superparticle in the
light--cone notation acquires the form \footnote{ In what follows we use
the following realization for $D=6$ ``Pauli" matrices ${\g}^{\u M}_{{\u
\mu}{\u \nu}}$ and ${\tilde \g}^{{\u M}{{\u \mu}{\u \nu}}}$ :  $$ {\g}^{\u
0}_{{{\u \mu}}{{\u \nu}}}=\left( \begin{array}{cc} -{\epsilon}_{AB} & 0 \\
0 & -{\epsilon}_{{\dot A}{\dot B}} \\ \end{array} \right), \quad {\tilde
\g}^{{\u 0}{\u \mu}{\u \nu}}=\left( \begin{array}{cc} {\epsilon}^{AB}& 0
\\ 0 & {\epsilon}^{{\dot A}{\dot B}} \\ \end{array} \right), \quad $$ \be
\label{507} {\g}^{\u 5}_{{{\u \mu}}{{\u \nu}}}=\left( \begin{array}{cc}
-{\epsilon}_{AB} & 0 \\ 0 & {\epsilon}_{{\dot A}{\dot B}} \\ \end{array}
\right), \quad {\tilde \g}^{{\u 5}{\u \mu}{\u \nu}}=\left(
\begin{array}{cc} -{\epsilon}^{AB}& 0 \\ 0 & {\epsilon}^{{\dot A}{\dot B}}
\\ \end{array} \right), \quad \ee $$ {\g}^{i}_{{\u \mu}{\u \nu}}=\left(
\begin{array}{cc} 0 & {\s}^{i}_{A{\dot A}} \\ {\tilde \s}^i_{{\dot A}A} &
0 \\ \end{array} \right), \quad {\bar \g}^{i{\u \mu}{\u \nu}}=\left(
\begin{array}{cc} 0 & -{\s}^{{i}{A{\dot A}}}\\ -{\tilde \s}^{i{\dot A}A} &
0 \\ \end{array} \right), \quad $$ and $2\times 2$ matrices
${\s}^i_{A{\dot A}}$ and ${\bar \s}^{i{\dot A}A}$ are defined above} $$
{\Pi}^{++}_{IIA}\equiv dX^{{++}}+2id{\Th}^{(1)+{A{\dot
I}}}{\Th}^{(1)+}_{A{\dot I}}+ 2id{\Th}^{(2)+{\dot
A}{I}}{\Th}^{(2)+}_{{\dot A}I}= e_{\tau}{\Pi}^{++}_{{\tau}IIA}, $$ \be
\label{509} {\Pi}^{--}_{IIA}\equiv dX^{{--}}+2id{\Th}^{(1)-{{\dot A}{\dot
I}}}{\Th}^{(1)-}_{{\dot A}{\dot I}}+
2id{\Th}^{(2)-{A}{I}}{\Th}^{(2)-}_{{A}I}= e_{\tau}{\Pi}^{--}_{{\tau}IIA}
\ee $$ {\Pi}^{i}_{IIA}\equiv dX^{{i}}-id{\Th}^{(1)+{{A}{\dot
I}}}{\s}^i_{A{\dot A}}{\Th}^{(1)-{\dot A}}_{~~~~{\dot I}}-
id{\Th}^{(1)-{{\dot A}{\dot I}}}{\bar \s}^i_{{\dot
A}A}{\Th}^{(1)+{A}}_{~~~~{\dot I}}+ $$ $$
d{\Th}^{(2)-{{A}{I}}}{\s}^i_{A{\dot A}}{\Th}^{(2)+{\dot A}}_{~~~~{I}}+
id{\Th}^{(2)+{{\dot A}{I}}}{\bar \s}^i_{{\dot
A}A}{\Th}^{(2)-{A}}_{~~~~{I}}= e_{\tau}{\Pi}^{i}_{{\tau}IIA}.  $$

     In particular, $D_{   {\hat q}}{\Th}^{+}$ satisfy the constraints $$
(D_{   {\hat p}}{\Th}^{(1)+{\dot I}}_{~~~~~A}) (D_{   {\hat
q}}{\Th}^{(1)+~A}_{~~~~{\dot I}})+ (D_{   {\hat
p}}{\Th}^{(2)+{I}}_{~~~~~{\dot A}}) (D_{   {\hat q}}{\Th}^{(2)+~{\dot
A}}_{~~~~{I}})= -{\d}_{{\hat p}{\hat q}}{\Pi}^{++}_{{\tau}IIA}, $$ which
follow from \p{509} and imply that the number of independent fields in the
$8\times 8$ matrices \be \label{5066} (D_{   {\hat q}}{\Th}^{(1)+{\dot
I}}_{~~~~~A},~ D_{   {\hat q}}{\Th}^{(2)+{I}}_{~~~~~{\dot A}}) \ee is
equal to the number of independent parameters of the local worldline
$SO(1,1)\times SO(8)$ symmetry, transforming these matrices.  Really, the
action \p{505} was obtained by dimensional reduction of the $D=10 ~N=1$
superparticle action \p{501} \cite{gs92}, in which the components of
$D_{\hat q}{\Th}^{\hat{\u \a}}$ form matrices of $SO(1,1)\times SO(8)$
when ${\Th}^{\hat{\u \a}}$ is taken in the light--cone basis:
${\Th}^{\hat{\u \a}}=({\Th}^{+{\hat Q}},{\Th}^{-{\dot{\hat Q}}})$.  The
reduction prescription applied in this paper uses these matrices with
indices ${\hat{\u \a}}$ and ${\hat Q},~{\dot{\hat Q}}$ decomposed in
accordance with the $D=6$ spinor structure.

      The fields ${\Th}$ being the scalar worldline superfields, the
transformation law for \p{5066} is determined by the transformation
properties of the fermionic covariant derivatives \p{209}.  The
transformation matrices $D'_{   {\hat q}}{\eta}_{\hat q}$ are restricted
by \p{208}, and, as a consequence, by \be \label{5067} (D'_{   {\hat
p}}{\eta}_{\hat s}) (D'_{   {\hat q}}{\eta}_{\hat s})={1\over
8}{\d}_{{\hat p}{\hat q}} (D'_{   {\hat t}}{\eta}_{\hat s}) (D'_{   {\hat
t}}{\eta}_{\hat s}) \ee

Hence, they take their values in the $SO(1,1)\times SO(8)$ subgroup of the
$n=8$ local worldline supersymmetry group. One can use this symmetry to
gauge away all the fields in \p{5066} \cite{gs92}.  \footnote{Moreover,
the complete worldline supersymmetry can be used to fix a gauge $$
({\Th}^{(1)+{\dot I}}_{A}, {\Th}^{(2)+{I}}_{\dot A})={\eta}^{\hat q}, $$
where all the world--line supersymmetries are broken. This guarantees the
possibility of gauge fixing \p{512}.} Fixing the gauge, one should take
into account that either $(D_{   {\hat q}}{\Th}^{(1)+{\dot I}}_{~~~~~A},~
D_{   {\hat q}}{\Th}^{(2)+{I}}_{~~~~~{\dot A}})$ or $(D_{   {\hat
q}}{\Th}^{(1)-{\dot I}}_{~~~~~{\dot A}},~ D_{   {\hat
q}}{\Th}^{(2)-{I}}_{~~~~~{A}})$ should have a non--vanishing determinant.
In what follows we will consider the first matrix to be non--degenerate,
so only the gauges compatible with this requirement are admissible. For
example, we can fix the matrix in question to be the unit one.
Decomposing the $SO(8)$--index ${\hat q}$ into three indices $SU(2)\times
SU(2)\times SO(2):~{\hat q}\rightarrow ({\tilde A},{\dot{\tilde A}},{\hat
A})$, we write the gauge fixing conditions as \be \label{512} D^{\hat
A}_{{    }{\tilde A}{\dot{\tilde A}}}{\Th}^{(1)+{\dot I}}_{~~~~~A}=
{\d}^{(1){\hat A}}{\d}^{\dot I}_{\dot A}{\epsilon}_{{\tilde A}A};\quad
~D^{\hat A}_{{    }{\tilde A}{\dot{\tilde A}}}{\Th}^{(2)+{I}}_{~~~~~{\dot
A}}= {\d}^{(2){\hat A}}{\d}^{I}_{A}{\epsilon}_{{\dot{\tilde A}}{\dot A}};
\ee

      The general solution to the  geometrodynamical constraint  \p{509}
      can be written in the gauge \p{512} in the following form:  $$
D^{\hat 1}_{{    }{\tilde A}{\dot{\tilde A}}}{\Th}^{(1)-{\dot
I}}_{~~~~~{\dot A}}= -{\bar \s}^j_{{\dot A}{\tilde A}}{F}^j{\d}^{\dot
I}_{\dot A};\quad D^{\hat 2}_{{    }{\tilde A}{\dot{\tilde
A}}}{\Th}^{(1)-{\dot I}}_{~~~~~{\dot A}}= -{\bar \s}^{j{\dot
I}}_{~~{\tilde A}}{E}^j{\epsilon}_{{\dot A}{\dot{\tilde A}}}; $$ \be
\label{513} D^{\hat 1}_{{    }{\tilde A}{\dot{\tilde
A}}}{\Th}^{(2)-{I}}_{~~~~~{A}}= {\epsilon}_{A{\tilde A}}{E}^j{\bar
\s}^{jI}_{\dot{\tilde A}};\quad D^{\hat 2}_{{    }{\tilde A}{\dot{\tilde
A}}}{\Th}^{(2)-{I}}_{~~~~~{A}}= -{\d}^I_{{\tilde A}}{F}^j{\bar
\s}^{j}_{\dot{\tilde A}{A}};~ \ee $$ {\Pi}^{++}_{\tau}=2;\qquad
{\Pi}^{--}_{\tau}={E}^i{E}^i+{F}^i{F}^i;\qquad {\Pi}^i_{\tau}={F}^i; $$
Higher components of the superfields ${E}^i$ and ${F}^i$ are expressed in
terms of the leading components of ${\Th}^{(1)}$ and ${\Th}^{(2)}$:  $$
D^{\hat 1}_{{    }{\tilde A}{\dot{\tilde A}}}{E}^i=
-i{\s}^i_{I{\dot{\tilde A}}} {{{\partial}{\Th}^{(2)-I}_{\tilde
A}}\over{{\partial}{\tau}}};\qquad D^{\hat 2}_{{    }{\tilde
A}{\dot{\tilde A}}}{E}^i= -i{\bar \s}^{i{\dot I}}_{~~{\tilde A}}
{{{\partial}{\Th}^{(1)-}_{{\dot{\tilde A}}{\dot
I}}}\over{{\partial}{\tau}}};~ $$ $$ D^{\hat 1}_{{    }{\tilde
A}{\dot{\tilde A}}}{F}^i= i{\bar \s}^i_{{\dot{A}}{\tilde A}}
{{{\partial}{\Th}^{(1)-{\dot A}}_{\dot{\tilde
A}}}\over{{\partial}{\tau}}};\qquad D^{\hat 2}_{{    }{\tilde
A}{\dot{\tilde A}}}{F}^i= i{\bar \s}^{i}_{A{\dot{\tilde A}}}
{{{\partial}{\Th}^{(2)-A}_{{{\tilde A}}}}\over{{\partial}{\tau}}};~ $$ and
no dynamical equations appear.

         One can see that the Type $IIA$  geometrodynamical constraint
	 does not put the theory on the mass shell and thus the Lagrange
multiplier $P^{\u M}_{{\hat q}}$ does not contain redundant propagating
degrees of freedom. The superfield equations, which one can obtain by
varying the action \p{505} with respect to the Lagrange multiplier
$Q_{{\hat p}{\hat q}}$, are consequences of the geometrodynamical
constraint \p{415} and produce no additional restrictions on the fields.
So, $Q_{{\hat p}{\hat q}}$ is a purely auxiliary degree of freedom, which
can be eliminated (as well as auxiliary fields in $P^{\u M}_{{\hat q}}$)
with the help of the gauge symmetries \p{5065} and \p{506} of the action
\p{505}.  Hence, this action consistently describes $D=6$ Type $IIA$
superparticle dynamics.

	The different situation is in the case of a  $IIB ~D=6$
	superparticle.  Now the geometrodynamical constraint takes the
form $$ {\Pi}^{\u M}_{IIB}\equiv dX^{{\u M}}- id{\Th}^{({\hat A}){{\u
\mu}{\dot I}}}{\g}^{{\u M}}_{{{\u \mu}}{{\u \nu}}} {\Th}^{({\hat A}){\u
\nu}}_{\dot I}= e_{\tau}{\Pi}^{\u M}_{{\tau}IIB}, $$ where $({\hat A})$ is
the $SO(2)$ index of $N=2$ target space supersymmetry.  Its light--cone
decomposition can be written as follows $$ {\Pi}^{++}_{IIB}\equiv
dX^{{++}}+2id{\Th}^{({\hat A})+{A{\dot I}}}{\Th}^{({\hat A})+}_{A{\dot
I}}= e_{\tau}{\Pi}^{++}_{{\tau}IIB}, $$ \be \label{510}
{\Pi}^{--}_{IIB}\equiv dX^{{--}}+2id{\Th}^{({\hat A})-{{\dot A}{\dot
I}}}{\Th}^{({\hat A})-}_{{\dot A}{\dot I}}=
e_{\tau}{\Pi}^{--}_{{\tau}IIB}, \ee $$ {\Pi}^{i}_{IIB}\equiv
dX^{{i}}-id{\Th}^{({\hat A})+{{A}{\dot I}}}{\s}^i_{A{\dot A}}{\Th}^{({\hat
A})-{\dot A}}_{~~~~{\dot I}}- id{\Th}^{({\hat A})-{{\dot A}{\dot I}}}{\bar
\s}^i_{{\dot A}A}{\Th}^{({\hat A})+{A}}_{~~~~{\dot I}}=
e_{\tau}{\Pi}^{i}_{{\tau}IIB}.  $$ In \p{510} we decomposed the
$SU^{*}(4)$ spinor index into a pair of $SU(2)\times SU(2)$ indices and
introduced $SO(1,1)$ indices $+$ and $-$ as in the $IIA$ case:  $$
{\Th}^{({\hat A}){\u \mu}}= ({\Th}^{({\hat A})+A{\dot I}},{\Th}^{({\hat
A})-{\dot A}{\dot I}}); $$ Following the same arguments that have been
used in the analysis of the $IIA$ geometrodynamical constraint , we can
 gauge fix the matrix $(D_{   {\hat q}}{\Th}^{({\hat A})+{\dot
I}}_{~~~~~A})$ to be the unit matrix $$ D^{\hat A}_{{    }{\tilde
A}{\dot{\tilde A}}}{\Th}^{({\hat A})+{\dot I}}_{~~~~~A}= {\d}^{({\hat
A}){\hat A}}{\d}^{\dot I}_{\dot A}{\epsilon}_{{\tilde A}A}.  $$ In this
gauge the general solution to the $IIB$ geometrodynamical constraint is $$
D^{\hat A}_{{    }{\tilde A}{\dot{\tilde A}}}{\Th}^{(A)-{\dot
A}}_{~~~~{\dot I}}= {\d}^{{\hat A}({\hat A})}{\bar \s}^{j{\dot
A}}_{~~{\tilde A}}{G}^j{\epsilon}_{{\dot{\tilde A}}{\dot I}};~ $$ \be
\label{514} {\Pi}^{++}_{\tau}=2;\qquad {\Pi}^{--}_{\tau}={G}^i{G}^i;\qquad
{\Pi}^i_{\tau}=4{G}^i; \ee One can see that this solution contains twice
less independent fields in comparison with the $IIA$ case. So the
equations \p{510} are more restrictive than the constraints \p{509}. This
results in the mass shell equations arising from the higher order
selfconsistency conditions of \p{510}.  Hitting \p{514} by Grassmann
covariant derivatives and using their algebra $$ \{ D^{\hat A}_{
{\tilde A}{\dot{\tilde A}}},D^{\hat B}_{    {\tilde B} {\dot{\tilde B}}}\}
=2i{\epsilon}_{{\tilde A}{\tilde B}} {\epsilon}_{{\dot{\tilde
A}}{\dot{\tilde B}}}{\d}^{{\hat A}{\hat B}}
{{\partial}\over{{\partial}{\tau}}}, $$ we obtain the equations of motion
of the $D=6$ $IIB$ superparticle in the light--cone gauge:  $$
{{{\partial}{\Th}^{(1)-}_{{\dot{\tilde A}}{\dot
I}}}\over{{\partial}{\tau}}}=0.  $$

	This means that if we wrote down a worldline action for the Type
$IIB$ superparticle in the form similar to \p{505}, the Lagrange
multiplier $P_{\hat q}^{\u M}$ would contain redundant propagating degrees
of freedom and the classical dynamics of such a model would not correspond
to the dynamics of the $D=6$ $IIB$ Casalbuoni--Brink--Schwarz
superparticle.

\section{Conclusion}

       In this paper the doubly supersymmetric actions have been
       constructed for $N=2$ superparticles in  $D=3,4$ and for the Type
$IIA$ superparticle in $D=6$. These actions have been obtained by the
dimensional reduction of the superfield actions for $N=1$ superparticles
\cite{gs92} in $D>3$. They possess an explicit $n=2(D-2)$ local worldline
supersymmetry, replacing $\kappa$ -- symmetries of the corresponding
Casalbuoni--Brink--Schwarz superparticles which they classically
equivalent to. By comparison with the earlier constructed superfield
actions for $N=1$ \cite{stv}--\cite{gs92} and $N=2 ~D=3$ \cite{gs2}
superparticles the proposed actions contain additional Lagrange multiplier
terms, ensuring their invariance under the local superfield
transformations of the Lagrange multipliers in the actions.  They are
crucial for elimination of auxiliary fields from the Lagrange multipliers.

One may also apply the dimensional reduction procedure to the worldsheet
superfield actions of $N=1$ superstrings \cite{hsstr} to get superfield
actions for superstrings with extended supersymmetries in $D=4$ and $6$.
The doubly supersymmetric description of such models is of interest, in
particular, because of a connection, recently found between the equations
of the doubly supersymmetric formulation of $D=3~N=2$ Green--Schwarz
superstring and some exactly solvable models \cite{stvt}. For instance, it
was demonstrated that these equations are related to the equations of a
properly constrained supersymmetric WZNW model based on the $sl(2,{\bf
R})$ algebra. The study of the dynamics of $N=2$ superstrings in $D>3$ in
the doubly supersymmetric approach may reveal their relation with more
complicated WZNW models with extended two--dimensional supersymmetry,
based on the algebras $sl(2,{\bf C})$, $spin(1,5)$ and $spin(1,9)$.

{\bf Acknowledgments.} D.S. would like to thank Emery Sokatchev for
discussion and Nathan Berkovits for a comment. This work was partially
supported by grants of the Ministry of Science and Technology of Ukraine
and the INTAS Grants N 93--493--ext, N 94--2317 and N 96--0308.

\end{document}